\documentclass[pra,twocolumn,preprintnumbers,superscriptaddress]{revtex4} 

\usepackage{graphicx}
\usepackage{bm}
\usepackage{amsmath}
\usepackage{amssymb}
\usepackage{amsfonts}
\usepackage{fancyhdr}
\usepackage{slashed}
\usepackage{latexsym,epsfig,bbm}

\newcommand{\md}{\mathrm{d}}
\newcommand{\bra}[1]{\langle{#1}|}
\newcommand{\ket}[1]{|{#1}\rangle}

\newcommand{\td}[2]{\frac{\md #1}{\md #2}}

\newcommand{\realsum}{\displaystyle\sum}

\newcommand{\figref}[1]{Fig.~\ref{#1}}

\def \d {\mathrm{d}}

\usepackage{color}
\definecolor{dblue}{rgb}{.2,.2,.8}

\definecolor{tom}{rgb}{.8,.4,.1}

\begin{document}

\title{Nondestructive Probing of Means, Variances, and Correlations of Ultracold-Atomic-System Densities via Qubit Impurities}

\author{T. J. Elliott}
\email{thomas.elliott@physics.ox.ac.uk}
\affiliation{Department of Physics, Clarendon Laboratory, University of Oxford, Parks Road, Oxford OX1 3PU, United Kingdom}
\author{T. H. Johnson}
\affiliation{Centre for Quantum Technologies, National University of Singapore, 3 Science Drive 2, 117543, Singapore}
\affiliation{Department of Physics, Clarendon Laboratory, University of Oxford, Parks Road, Oxford OX1 3PU, United Kingdom}
\affiliation{Keble College, University of Oxford, Parks Road, Oxford OX1 3PG, United Kingdom}

\date{\today}

\begin{abstract}
We show how impurity atoms can measure moments of ultracold atomic gas densities, using the example of bosons in a one-dimensional lattice. This builds on a body of work regarding the probing of systems by measuring the dephasing of an immersed qubit. We show this dephasing is captured by a function resembling characteristic functions of probability theory, of which the derivatives at short times reveal moments of the system operator to which the qubit couples. For a qubit formed by an impurity atom, in a system of ultracold atoms, this operator can be the density of the system at the location of the impurity, and thus, means, variances, and correlations of the atomic densities are accessible. 
\end{abstract}
\maketitle 

\section{Introduction}
The impressive quantum control and versatility achieved in cold quantum gas experiments \cite{bloch2008} exemplifies their use for quantum technologies, such as quantum simulation \cite{lewenstein2012ultracold, johnson2014} and quantum information processing \cite{monroe2002}. Typical methods for measuring these systems, such as time-of-flight imaging of the momentum distribution~\cite{hadzibabic2003, altman2004, folling2005, bloch2008,perrin2009} and in-situ imaging of density~\cite{nelson2007,bakr2009,sherson2010,weitenberg2011}, require destroying the system. Each measurement is of a single-shot of the system state, and hence finding averages requires repeated trapping and cooling of the system.

In this work, we propose an alternative method where trapped atomic impurities forming qubit probes are immersed in the system, entangling system and probes. Each qubit is dephased, and measuring this dephasing reveals information about the system \cite{recati2005, bruderer2006, sabin2014, hangleiter2015, johnson2015, cosco2015}. The dephasing relates to the system operator to which the probes couple, and for typical low-energy interactions between cold atoms, this is the density of the system at the location of the impurity. We show that this enables measurement of not only the mean density, but also density variances and correlations between different locations. This exploits the similarity of the dephasing function to characteristic functions of probability theory, allowing moments of the density to be determined. A major advantage of this scheme is that it is potentially non-destructive, leaving the system intact and allowing for repeated measurements.

As our example, we consider a gas of bosons in a one-dimensional lattice, described by the Bose-Hubbard model. We demonstrate, by simulating the gas and impurity atoms for typically accessible experimental parameters, that our protocol can faithfully capture, even in the presence of errors, the behavior of density-related properties, over a range of phases of the system, thus providing an additional tool for measuring ultracold atoms. Moreover, since we develop the protocol generically, independent of the particular system, our work contributes to and furthers the ongoing development of methods for characterizing systems using coupled non-equilibrium dynamics, in cold atom and other setups, by observing the dephasing of a qubit in an environment~\cite{goold2011, knap2012, borrelli2013, haikka2013,  dorner2013, mazzola2013, batalhao2014, haikka2014}.

\section{Generic Qubit Probe}
\subsection{Extracting the Dephasing Function}
We consider a qubit with Hamiltonian $H_q=\omega_q\ket{1}\bra{1}$ immersed in a system with Hamiltonian $H_{\mathrm{sys}}$. Here $\omega_q$ is the energy difference between the qubit states $\{\ket{0},\ket{1}\}$. State $\ket{1}$ couples with strength $\kappa$ to system operator $H_{\mathrm{int}}$, which we call the interaction Hamiltonian, such that the combined system is described by the Hamiltonian
\begin{equation}
\label{fullhamiltonian}
\mathcal{H}=\mathbb{I}\otimes H_{\mathrm{sys}} + H_q\otimes\mathbb{I}+\kappa\ket{1}\bra{1}\otimes H_{\mathrm{int}},
\end{equation}
where $\mathbb{I}$ is the identity operator. We define the Hamiltonians $H_0= H_{\mathrm{sys}}$ and $H_1=H_{\mathrm{sys}}+\kappa H_{\mathrm{int}}$, describing the system evolution for each qubit state~\footnote{More generally, we can define $H_j= \bra{j} \mathcal{H} \ket{j}$ for $j=0,1$. Additionally, while here we have $H_0= H_{\mathrm{sys}}$ and $H_1=H_{\mathrm{sys}}+\kappa H_{\mathrm{int}}$, the crucial quantity in our analysis is $H_1 - H_0,$ here equal to $\kappa H_{\mathrm{int}}$. Thus, the state $\ket{0}$ can also be allowed to interact with the system, provided $H_1-H_0$ has the desired form.}.

\begin{figure}
\includegraphics[width=\linewidth]{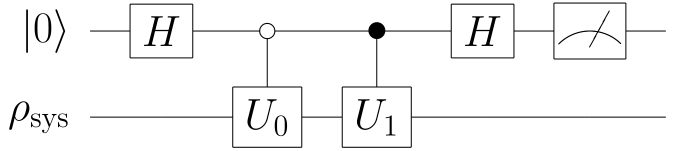}
\caption{Quantum circuit for obtaining the dephasing function from measurements of the probe qubit. Here, $H=(\sigma_x+\sigma_z)/\sqrt{2}$ is the Hadamard gate, and $U_j=e^{-iH_jt}$ is the evolution of the system under Hamiltonians $H_j$ for a time $t$.}
\label{figcircuit}
\end{figure}

In a Ramsey-like scheme [\figref{figcircuit}], we initialize the qubit at times $t<0$ in the non-interacting state $\ket{0}$ and the system in state $\rho_{\mathrm{sys}}$. At $t=0$, the qubit is suddenly switched to the state $(\ket{0}+\ket{1})/\sqrt{2}$~\footnote{Equivalently, the qubit can be initialized in the $\ket{+}$ state with $\kappa=0$ for $t<0$, with $\kappa$ then ramped up quickly at $t=0$.}. For times $t\geq0$, the system and qubit then evolve according to the Hamiltonian \eqref{fullhamiltonian}, such that at time $t$ the qubit state is
$$\rho_q(t)=\frac{1}{2}\begin{pmatrix}1&e^{i\omega_qt}L(t)\\e^{-i\omega_qt}L^*(t)&1\end{pmatrix}.$$
Here we define the dephasing function (sometimes called the overlap function) $L(t)=\langle e^{iH_1t}e^{-iH_0t}\rangle$, with expectations taken for the initial system state $\rho_{\mathrm{sys}}$.  Values of the dephasing function $L(t)$ are related to the expectations of the Pauli operators: $\langle\sigma_x\rangle=\Re(e^{i\omega_qt}L(t))$ and $\langle\sigma_y\rangle=\Im(e^{i\omega_qt}L(t))$.

The dephasing function has been investigated previously to study properties such as the orthogonality catastrophe in ultracold fermions \cite{goold2011,knap2012}, the Luttinger parameter \cite{recati2005}, and superfluid excitations \cite{cosco2015}, as a method of thermometry~\cite{bruderer2006,sabin2014,hangleiter2015,johnson2015}, and to extract work distributions \cite{dorner2013,mazzola2013,batalhao2014}. In this work, we proceed by investigating the derivatives of the dephasing function, motivated by the similar structure it shares with the characteristic function of a probability distribution \cite{fusco2014}.

\subsection{Derivatives of the Dephasing Function}
In a strong-coupling limit, where energies associated with the interaction Hamiltonian $\kappa H_{\mathrm{int}}$ are much larger than those of the system Hamiltonian, at sufficiently short times that $\langle e^{iH_\mathrm{sys} t}\rangle\approx1$ the dephasing function tends towards $L_{\mathrm{strong}}(t) = \langle e^{i\kappa H_{\mathrm{int}} t}\rangle$. This is the characteristic function of the interaction Hamiltonian. Thus, like such functions in probability theory, from this all moments of $H_{\mathrm{int}}$ can be obtained directly from derivatives;
$$\langle H_{\mathrm{int}}^n\rangle=\frac{1}{(i\kappa)^n}\left.\frac{\d ^n L_{\mathrm{strong}}(t)}{\d t^n}\right|_{t=0}.$$
The drawback of this limit is that it may not be easily accessible in experiment for some systems, as it requires a large coupling strength to be engineered between system and probes. Increasing the coupling strength also decreases the timescales over which $L(t)$ evolves, and hence requires sufficiently quick switching of the qubit state or ramping of $\kappa$, and sufficiently fine time resolution of the qubit measurements that the derivatives at $t=0$ are resolvable.

Positively, it is possible to extract the first two moments of $H_{\mathrm{int}}$ for arbitrary $\kappa$. Considering the derivatives of $L(t)$ directly:
\begin{equation}
\label{singlederivative}
\langle H_{\mathrm{int}}\rangle = \frac{1}{i\kappa}\left.\td{L(t)}{t}\right|_{t=0},
\end{equation}
and 
\begin{equation}
\label{twoderivative}
\langle H_{\mathrm{int}}^2\rangle = \frac{-1}{\kappa^2} \Re\left(\left.\frac{\d^2 L(t)}{\d t^2}\right|_{t=0}\right),
\end{equation}
where we have used that the commutator $[H_0,H_1]$ is anti-Hermitian and thus has a purely imaginary expectation value. We now focus on these first two moments of the interaction Hamiltonian $H_{\mathrm{int}}$, as they can be extracted from derivatives of $L(t)$ for any $\kappa$, allowing more experimental flexibility. 

\subsection{Estimation Protocol and Errors}
\label{secerror}
At short times, the first and second derivatives Eqs.~\eqref{singlederivative} and \eqref{twoderivative} dominate the imaginary and real parts of the dephasing function derivatives respectively, and thus it is possible to extract the first two moments $\langle H_{\mathrm{int}}\rangle$ and $\langle H_{\mathrm{int}}^2\rangle$ by fitting linear and quadratic functions to the initial behavior of $\Im(L(t))$ and $\Re(L(t))$ respectively, using that $L(0)=1$.
More precisely, we first obtain estimates $\bar{L}(n \Delta t)$ of values $L(n \Delta t)$ at discrete times $n\Delta t$, for integer $n$ up to a maximum $N_\epsilon$ corresponding to time $t_\epsilon = N_\epsilon \Delta t$. We then perform two least-squares estimations, minimizing
\begin{align}
\sum_{n=1}^{N_\epsilon} \left (\Im \left( \bar{L}(n \Delta t) \right) - \alpha n\Delta t \right )^2 , \nonumber \\
\sum_{n=1}^{N_\epsilon} \left (\Re \left( \bar{L}(n \Delta t) \right) -1 - \frac{\beta}{2} (n \Delta t)^2 \right )^2 , \nonumber
\end{align}
with respect to $\alpha$ and $\beta$, and take the minimizing values $\bar{\alpha}$ and $\bar{\beta}$ as our estimates of $\kappa \langle H_{\mathrm{int}}\rangle$ and $\kappa^2 \langle H_{\mathrm{int}}^2\rangle$ respectively \footnote{It is possible to further increase the accuracy of the fit by using an initial estimate of $L(n \Delta t)$ to estimate the variances of $\bar{L}(n \Delta t)$ and then weighting the least squares fit accordingly.}. The choice of $t_\epsilon$ should be such that it optimizes the number of points used in the fit without being so large that the imaginary and real parts of $L(t)$ stop behaving approximately linearly and quadratically respectively \footnote{Deviations from linear and quadratic behavior, respectively, could be accounted for by including higher-order terms in the fit, whose corresponding coefficients are discarded after the fit is made.}.

Each estimate $\bar{L}(n \Delta t)$ of $L(n \Delta t)$ is obtained by estimating $\langle\sigma_{\mu}\rangle$ ($\mu = \{x,y\}$), by averaging outcomes of $N$ repeated measurements of $\sigma_{\mu}$. The resulting estimate will be unbiased, and for enough measurements will be Gaussian with a variance that can be calculated from $L(n \Delta t)$, given by $(1+L)(1-L)/4N$, and hence attenuates to a few percent for a reasonable number of measurements. In our examples, we will treat this finite number of measurements as the main source of error in estimating $L(n \Delta t)$, though other noise-based errors (such as imprecision in the time at which measurements are made, and stochastic imperfections in the gate implementation) will behave in the same manner, so can be considered equivalently. We will also implicitly account for the error arising from the discrete and finite nature of the timesteps at which measurements are made, and in the backaction of the system-probe interaction on the system state. Other errors that we will not directly account for in our simulations are the finite time required to implement gates (which may be neglected when this occurs on timescales much shorter than that at which the measurements are made), and systematic imperfections in the gate implementation, which result in a deterministic multiplicative factor to the dephasing function \cite{johnson2015}, and may be accounted for by the corrective factor needed to ensure $L(0)=1$.

\section{Implementation in Ultracold Atomic Systems}
\label{secimplementation}

\begin{figure}
\includegraphics[width=\linewidth]{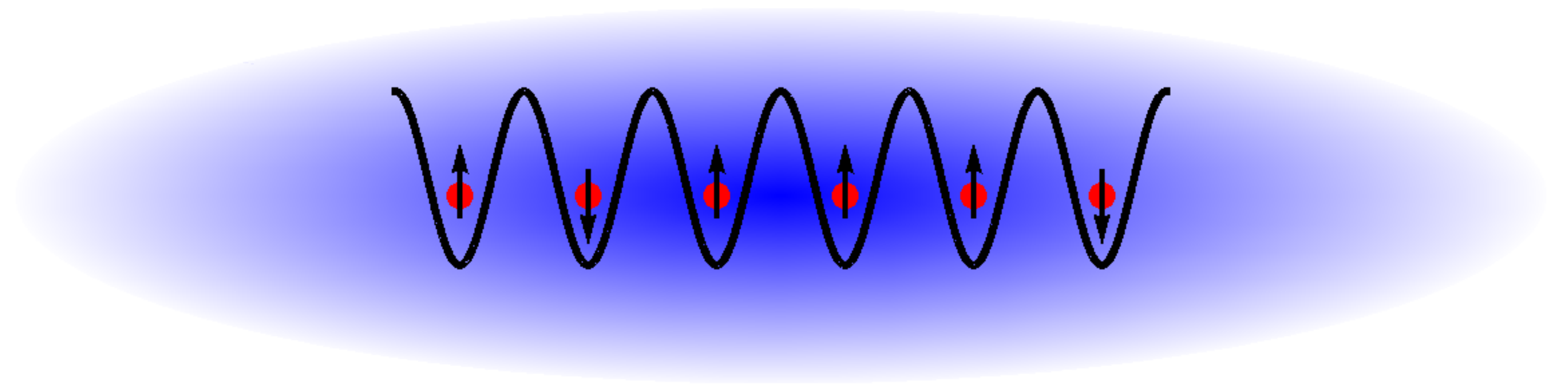}
\caption{Implementation of the protocol with ultracold atoms. The quantum gas is represented by the blue cloud, while the red atoms are the impurity probes. Moments of the interaction Hamiltonian are found through measurements of the impurities' internal states, represented by arrows.}
\label{figcoldatom}
\end{figure}

We now consider a possible implementation of the protocol for probing ultracold (single species) atomic gases. A probe qubit is formed by two internal states of an impurity atom of a different species \cite{knap2012}, trapped deeply such that its spatial wavefunction $\psi_q(\bm{x})$ is fixed \footnote{Thus, $\psi_q(\bm{x})$ may be treated as a c-number function. This approximation is valid when the
trapping potential is much stronger than the interaction energy $\kappa$.}. The necessary gates may be applied to the impurity qubit using a Rabi laser pulse, and measurements performed using gates and state-dependent fluorescing. With interparticle interactions suppressed within the impurity gas, multiple probes can be active simultaneously, as illustrated in \figref{figcoldatom}; such mixtures of quantum gases have already been achieved experimentally \cite{roati2002, inouye2004, gunter2006, zaccanti2006}. Repeat measurements do not require the gas and impurities to be retrapped, and can hence be non-destructive, though the measurement will perturb the system and is hence not non-demolition.

The impurity qubit and atoms comprising the system interact through $s$-wave scattering, potentially controlled through Feshbach resonances \cite{perrin2009,pethick2001bose}, hence we have an interaction of the form
$$H_{\mathrm{int}}=\int\d{\bm{x}}  n(\bm{x}) |\psi_q(\bm{x})|^2\equiv\tilde{n}(\bm{x}),$$
where $n(\bm{x})$ is the density of atoms in the system, and $\tilde{n}(\bm{x})$ is the system density course-grained over the impurity density. Our protocol thus allows probing of the moments of this course-grained density.  When the impurity is highly localized around $\bm{x}_i$, $\tilde{n}(\bm{x})$ is then given by the gas density at this point, $n(\bm{x}_i)$. In this case, our protocol will estimate the moments of the density at this point, $\langle n(\bm{x}_i)\rangle$ and $\langle n(\bm{x}_i)^2\rangle$ [Eqs.~\eqref{singlederivative} and \eqref{twoderivative}]. Additionally, since impurity atoms can be localized to regions smaller than the wavelength of light, the spatial resolution is potentially higher than for in-situ imaging. 

If the impurity is in a superposition of being localized to two distinct locations $\bm{x}_i$ and $\bm{x}_j$ (for example, by placing it in a lattice potential, and using tunneling as a beamsplitter operation \cite{daley2012}), then $\tilde{n}(\bm{x}) = (n(\bm{x}_i) + n(\bm{x}_j))/2$, and our protocol will then estimate $\langle n(\bm{x}_i) + n(\bm{x}_j) \rangle$ and $\langle (n(\bm{x}_i) + n(\bm{x}_j) )^2\rangle$. Using these and the previous results, an estimate for the correlation function $\langle n(\bm{x}_i)n(\bm{x}_j)\rangle$ is obtained. Alternatively, this can also be achieved using two qubits localized at $\bm{x}_i$ and $\bm{x}_j$ with entangled internal states $(\ket{00}+\ket{11})/\sqrt{2}$, thus behaving as a single effective qubit with a density equal to the sum of that of the individual probes.

\section{Simulation for Ultracold Atoms in a Lattice}
\label{secsimulation}

\begin{figure*}
\centering
\includegraphics[width=0.48\linewidth]{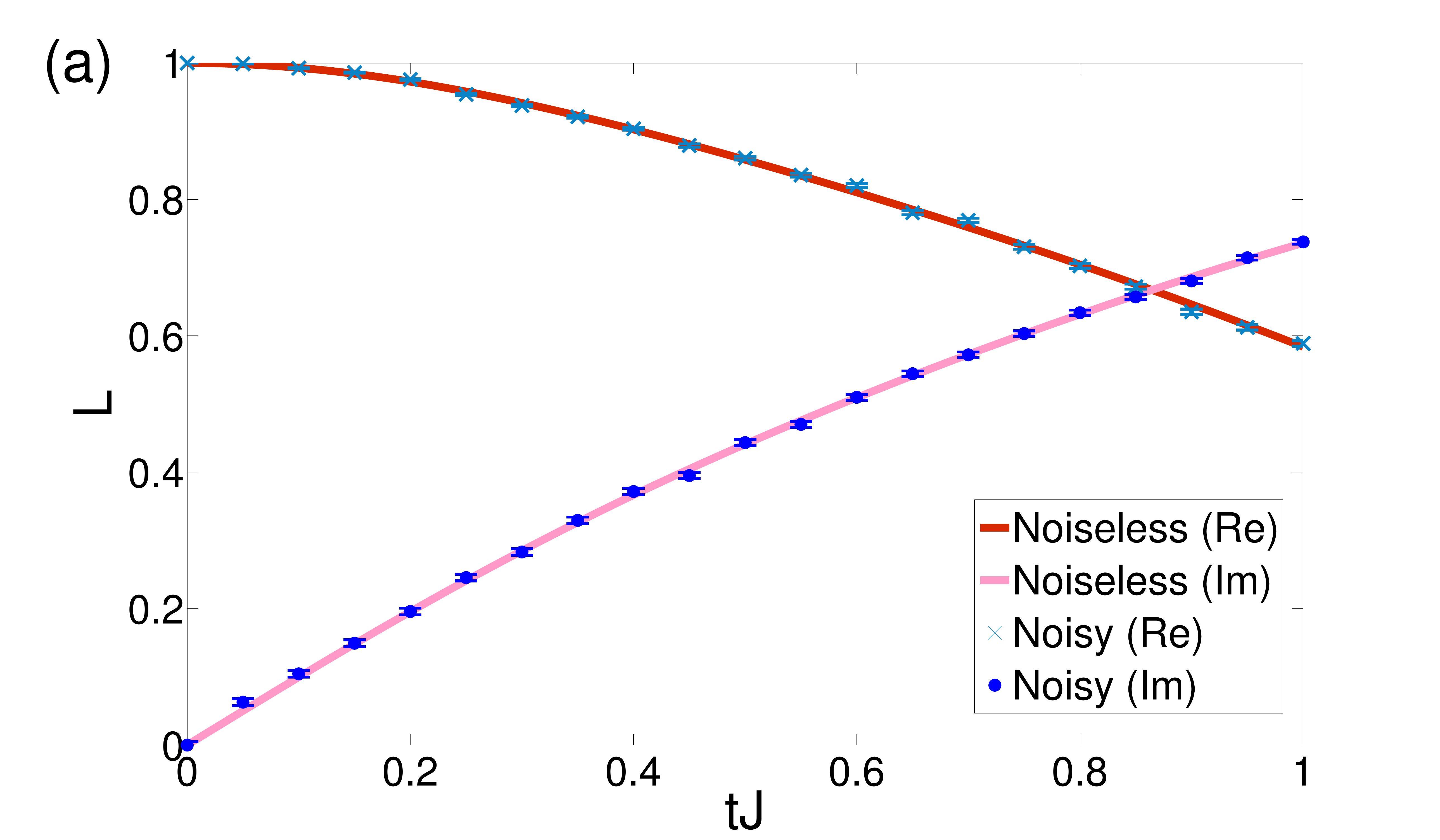}
\includegraphics[width=0.48\linewidth]{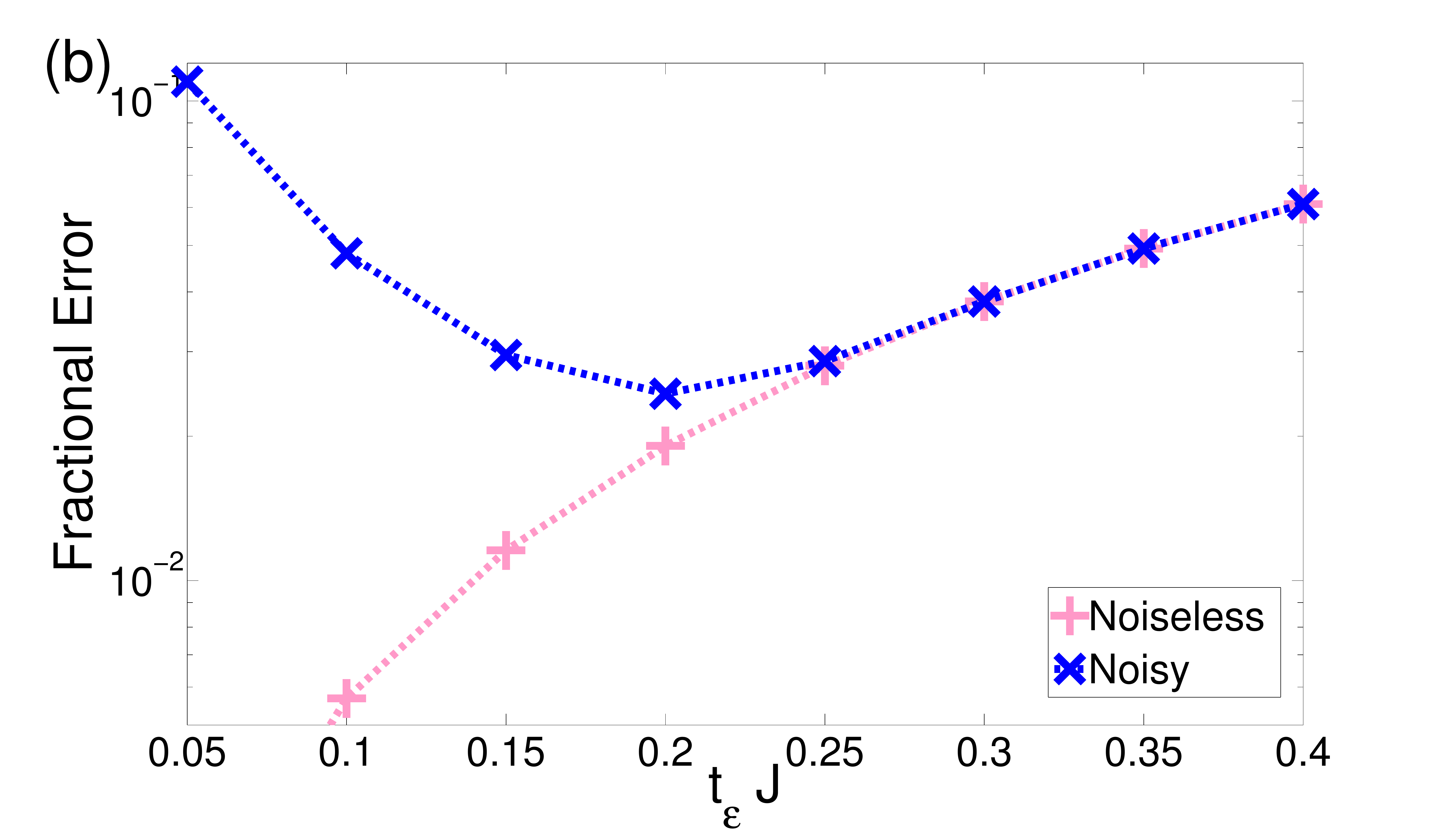}
\includegraphics[width=0.48\linewidth]{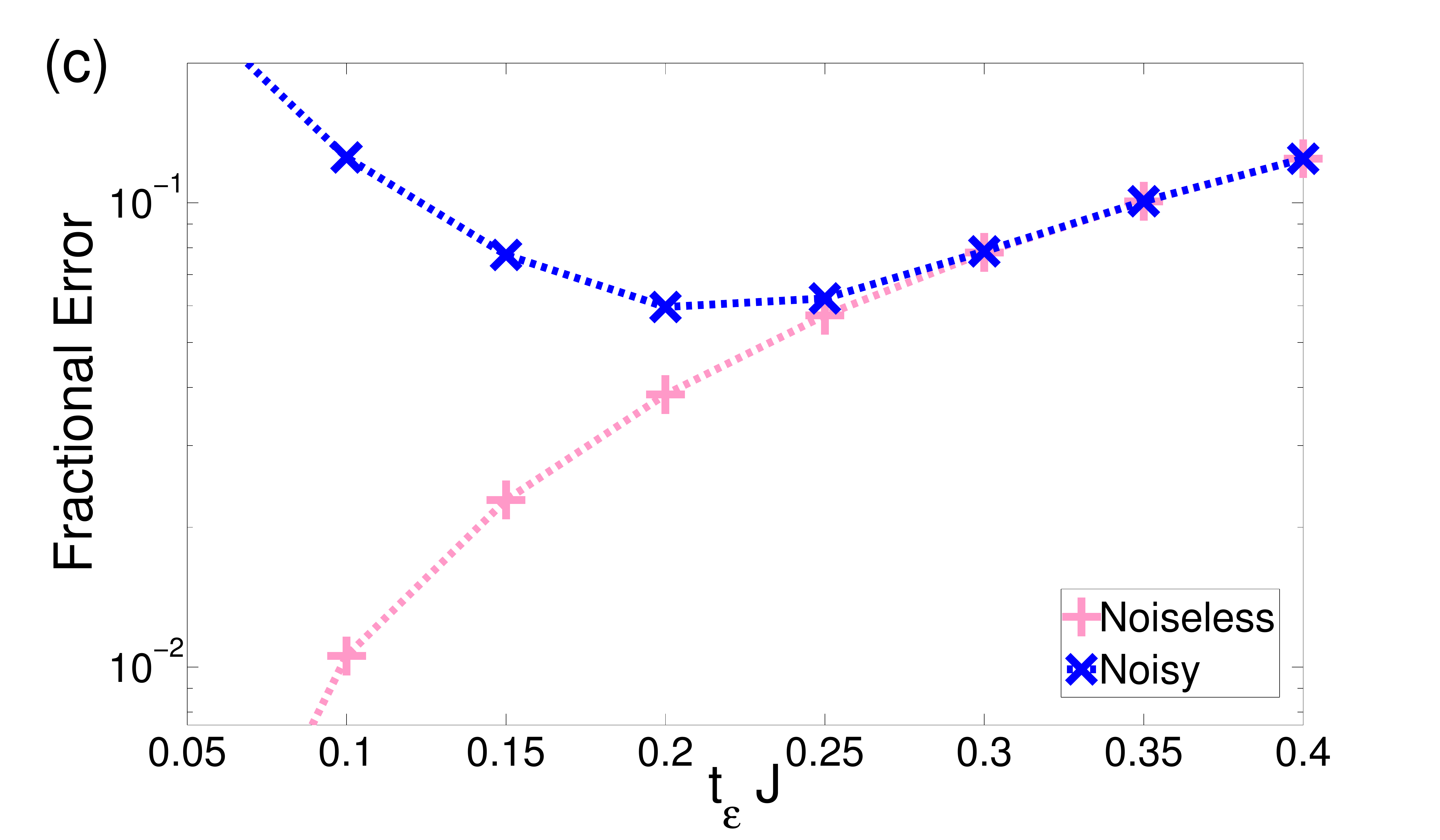}
\includegraphics[width=0.48\linewidth]{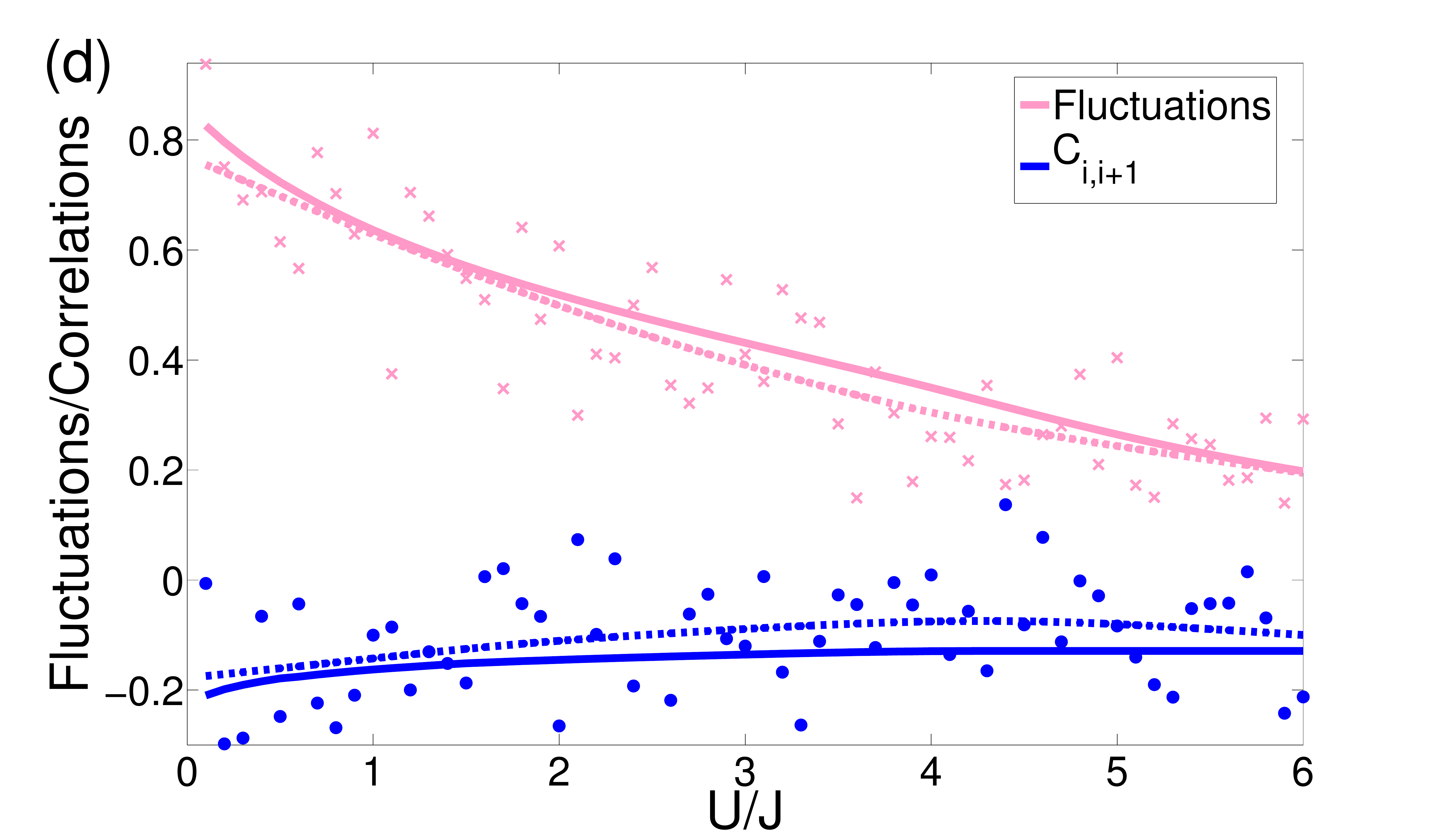}
\caption{Example application of the protocol: (a) The effect of Gaussian white noise on the dephasing function, with error bars denoting the standard deviation. The fractional error in the calculated (b) density and (c) square of density with increasing length of time used for the fit. (d) Fluctuations and correlations; solid lines show exact values, markers show the calculated values, and dashed lines the fits to calculated values. (a-c) simulate a system with $U=3J$, and for all plots $\kappa=J$ with a measurement resolution $\Delta tJ=0.05$.}
\label{figsimulation}
\end{figure*}

We now simulate an example application of the protocol to a Bose gas trapped in an optical lattice, obeying the one-dimensional Bose-Hubbard Hamiltonian \cite{jaksch1998,lewenstein2012ultracold}
\begin{equation}
\label{bhhamiltonian}
H_{\mathrm{sys}}=-J\realsum_{\langle i,j\rangle}(b^\dagger_ib_j+b^\dagger_jb_i)+U\realsum_ib^\dagger_ib^\dagger_ib_ib_i.
\end{equation}
Here $b_i$ ($b^\dagger_i$) annihilates (creates) a boson localized at site $i$ (with number operator $n_i=b^\dagger_i b_i$), and $\langle i,j\rangle$ denotes a sum over nearest-neighbor sites. The energies $J$ and $U$ parameterize hopping between neighboring sites and on-site interactions, respectively. We simulate a one-dimensional system with 101 lattice sites and unit filling factor, and we examine the system for $U/J$ in the interval $0.1:6$, spanning the whole phase diagram of the model at this filling factor \cite{lewenstein2012ultracold}. We are concerned entirely with investigating the ground state $\rho_\mathrm{sys}$ of the system.  

We use parameters that assume a realistic time-resolution in the measurements of the probe qubit dephasing. We choose $\kappa =J$, and $\Delta t=0.05J$, allowing us to take $N_\epsilon\lesssim20$ measurement points in our fit. For typical $J\sim10^3\hbar$Hz \cite{jaksch1998}, this corresponds to a measurement interval $\Delta t\sim 50\mu$s~\footnote{Specifically, parameters should be selected such that the relationship between timescales obeys $t_L>t_\epsilon>\Delta t>t_s$. Here $t_L \sim 1/ \kappa$ is the timescale at which $L(t)$ begins to deviate significantly from its $t=0$ behavior, and $t_s$ is timescale on which measurements and gates can be performed.}.

With these parameters, we simulate application of our protocol to the Bose-Hubbard model, by calculating $L(n \Delta t)$, adding stochastic noise to these values to simulate how estimates $\bar{L}(n \Delta t)$ would be obtained in a real experiment, then estimate one- and two-site correlations of site occupation numbers from these simulated values. These estimated values are then compared to exactly calculated equivalents. For both parts, calculating dephasing function $L(t)$ and ground state expectation values, we use the time-evolving block decimation (TEBD) algorithm \cite{vidal2004,johnson2011, TNT} (see Appendix \ref{sectebd} for details).

We begin our analysis by considering an impurity localized at some point $\bm{x}_i$ near the central site $i=51$ such that $H_{\mathrm{int}} = n_{i}$. We plot the real and imaginary parts of the dephasing function $L(t)$ for $U/J=3$ in \figref{figsimulation}(a), together with their simulated noisy estimates $\bar{L}(n \Delta t)$ obtained from $N=10^4$ measurements of the qubit. We apply our fitting procedure to estimate the derivatives of the dephasing function and thus the moments of the interaction Hamiltonian, giving estimates of $\langle n_{i} \rangle$ and $\langle n_{i}^2 \rangle$. In \figref{figsimulation}(b) we analyze the choice of $t_\epsilon=N_\epsilon\Delta t$ to use in the estimating the expected occupation $\langle n_{i} \rangle$, by averaging over many noisy trajectories. As expected, the accuracy initially increases with $N_\epsilon$, benefiting from an increase in the number of points used in the fit due to a corresponding decrease in random error. However, for larger $t_\epsilon$ the accuracy decreases with $N_\epsilon$ as times are included for which higher-order terms in $L(t)$ beyond the linear fit begin to play a significant role and thus introduce a systematic error. The same is shown for $\langle n_{i}^2 \rangle$ in \figref{figsimulation}(c), with the slight decrease in accuracy highlighting the increasing difficulty in estimating higher order derivatives and thus moments. We find $t_\epsilon J=0.2$ to be approximately optimal for estimating the moments. 

With this $t_\epsilon$ we assume the protocol is repeated with the impurities configured so as to probe $H_{\mathrm{int}} = n_{j}$ and $H_{\mathrm{int}} = n_{i} + n_{j}$. We then estimate the mean $\langle n_i \rangle$ and variance $\langle n_i^2 \rangle -\langle n_i \rangle^2$ of occupation at the central site, and the correlations $C_{i,j}=\langle n_i n_j \rangle - \langle n_i \rangle \langle n_j \rangle$ between this central site and neighboring site $j=i+1$. The results are plotted in \figref{figsimulation}(d). The markers show the value calculated from a single run of a noisy trajectory at each $U/J$, and the dashed lines a cubic smoothing spline fit \footnote{The cubic smoothing spline fit of a discrete set of data $y_i$ is defined as the function $f(x)$ that minimizes $\sum_i(y_i-f(x_i))^2-\lambda\int(\d^2 f/\d x^2)^2\d x$ \cite{hastie1990generalized}, where $\lambda$ is the `smoothing parameter'. We use $\lambda=10^4$ in order to obtain a very smooth fit.} over varying interaction strength for these data, while solid lines show the exact values. We observe that though individual data points display a noticeable error, fitting over the whole parameter range the quantitative values can be obtained faithfully to a high degree of accuracy.

\section{Summary and Outlook}
\label{secsummary}
We have investigated a method of using a qubit to probe properties of a system to which it is coupled. In particular, we have introduced a way to calculate moments of a so-called interaction Hamiltonian through the derivatives of a dephasing function that can be obtained through measurements of solely the qubit state, even for a weakly interacting qubit. Further, we have discussed how this protocol could be implemented for ultracold atomic systems, to reveal properties of the gas density.  We trialed the protocol by applying it successfully to simulations of atoms trapped in an optical lattice in one-dimension.

A direct application of this would be to probe phase transitions in the Bose-Hubbard model; for example, mean-field treatments, valid in higher dimensions, have found $\langle b_i \rangle$ to be a suitable order parameter \cite{rokhsar1991, sheshadri1993} for the superfluid-Mott insulator transition. The density fluctuations reveal whether this is non-zero, and hence also map out the phase diagram. Since our protocol is valid in any number of dimensions, this suggests that it could potentially be used as a way to non-destructively probe the transition. Our protocol could be extended beyond the current system, perhaps by manipulating the interaction between probe and system to allow other quantities to be probed, not just those related to density. For example, for bosons with spin, if $\kappa$ is sensitive to the spin (i.e. $\kappa\propto S_z$), then magnetization properties could also be probed. Alternatively, as the protocol has been derived generically, it could also be applied to other systems outside of cold atoms where similar probe-system interactions can be achieved, such as trapped ions \cite{dorner2013} and NMR spins \cite{batalhao2014}.

\section*{Acknowledgements}
The authors thank Stephen Clark and Dieter Jaksch for useful discussions. TJE thanks EPSRC (DTA) for financial support. THJ thanks the National Research Foundation and the Ministry of Education of Singapore for support. 

\appendix
\section{TEBD Algorithm}
\label{sectebd}
TEBD allows for the efficient classical simulation of the dynamical evolution of a pure quantum system on a one-dimensional lattice \cite{vidal2004,johnson2011, TNT}. The algorithm operates by storing the state as a matrix product state (MPS) \cite{verstraete2008}. An arbitrary quantum state may require an exponential scaling in the bond dimension $\chi$ of this MPS. However, ground states and low-lying states (including short time quenches from a ground state, as considered here) of local one-dimensional Hamiltonians may be very accurately represented by an MPS with a small bond dimension $\mathcal{O}(1-10^2)$ \cite{verstraete2006}, resulting in a tractable representation of the state. The MPS is then evolved for each discrete time step $\delta t$ according to a Hamiltonian $H$ formed of single-site and two-site nearest neighbor terms, after performing a Suzuki-Trotter decomposition \cite{suzuki1990} of the evolution operator $e^{-iH\delta t}$. The algorithm can also be used to find the ground state of a Hamiltonian, by performing the evolution in imaginary time. Expectation values of single-site and two-site operators can be calculated for an MPS after decomposing them into a matrix product representation.

The one-dimensional Bose-Hubbard model is well suited to such simulation, as the Hamiltonian \eqref{bhhamiltonian} consists only of one- and two-site nearest neighbor operators. We have here used imaginary time evolution to obtain the ground state MPS, and evolved this state according to each of $H_0$ and $H_1$. The overlap of these evolved states then gives us an expression for the dephasing function (up to a known phase). We also found the expectation values for the observables  $n_in_j$ $(n_i=b^\dagger_ib_i)$, for the purpose of providing exact values for comparison to the results obtained from the protocol. In our simulations we limited the maximum occupation per site to $4$ bosons, with $\chi=50$ and time step $\delta tJ=10^{-3}$. 

\bibliography{ref}

\end{document}